\documentclass[aps,prl,twocolumn,superscriptaddress]{revtex4-2}
\usepackage{bm}
\usepackage{graphicx}
\usepackage{dcolumn}
\usepackage{braket}
\usepackage{dsfont}
\usepackage{amsmath}
\usepackage{color}
\usepackage{amssymb}
\usepackage{amsthm}
\usepackage{amsmath}
\usepackage{qcircuit}
\usepackage{adjustbox}
\usepackage{tikz}
\usetikzlibrary{decorations.markings}
\usepackage{graphicx}
\usepackage{subfigure}
\usepackage{subcaption}
\usepackage{pgfplots}
\pgfplotsset{compat=1.3}
\usepgfplotslibrary{fillbetween}
\usetikzlibrary{patterns}
\usepackage[labelfont=bf,
   justification=Justified,
   format=plain]{caption}
\usepackage{mathtools}
\usepackage{xcolor}
\usepackage{enumerate}
\usepackage[many]{tcolorbox}    	
\usepackage{physics}
\usepackage{qcircuit}

\definecolor{amber}{rgb}{1.0, 0.75, 0.0}
\definecolor{color1}{RGB}{94, 201, 98}
\definecolor{color2}{RGB}{33, 145, 140}
\definecolor{color3}{RGB}{59, 82, 139}

\definecolor{plasma_color3}{RGB}{248, 149, 64}
\definecolor{plasma_color2}{RGB}{204, 71, 120}
\definecolor{plasma_color1}{RGB}{126, 3, 168}

\definecolor{colore}{rgb}{0.9, 0.9, 0.9}
\newtcolorbox{boxA}{
    colback = colore, 
    boxrule = 0pt  
}

\begin{document}

\title{
Universal emergence of local Zipf-Mandelbrot law
}

\author{Davide Cugini}
\affiliation{Dipartimento di Fisica, Universit\`a di Pavia, via Bassi 6, 27100,  Pavia, Italy}

\author{André Timpanaro}
\affiliation{Universidade Federal do ABC,  09210-580 Santo Andr\'e, Brazil}

\author{Giacomo Livan}
\affiliation{Dipartimento di Fisica, Universit\`a di Pavia, via Bassi 6, 27100,  Pavia, Italy}
\affiliation{Department of Computer Science, University College London, 66-72 Gower Street, WC1A 6EA London, UK}

\author{Giacomo Guarnieri}
\affiliation{Dipartimento di Fisica, Universit\`a di Pavia, via Bassi 6, 27100,  Pavia, Italy}

\begin{abstract}
A plethora of natural and socio-economic phenomena share a striking statistical regularity, that is the magnitude of elements decreases with a power law as a function of their position in a ranking of magnitude. Such regularity is known as  Zipf-Mandelbrot law (ZM), and plenty of problem-specific explanations for its emergence have been provided in different fields. Yet, an explanation for ZM ubiquity is currently lacking. 
In this paper we first provide an analytical expression for the cumulants of any ranked sample of i.i.d. random variables once sorted in decreasing order.
Then we make use of this result to rigorously demonstrate that, whenever a small fraction of such ranked dataset is considered, it becomes statistically indistinguishable from a ZM law. We finally validate our results against several relevant examples.
\end{abstract}

\begin{figure*}
\centering
\includegraphics[width=0.8\textwidth]{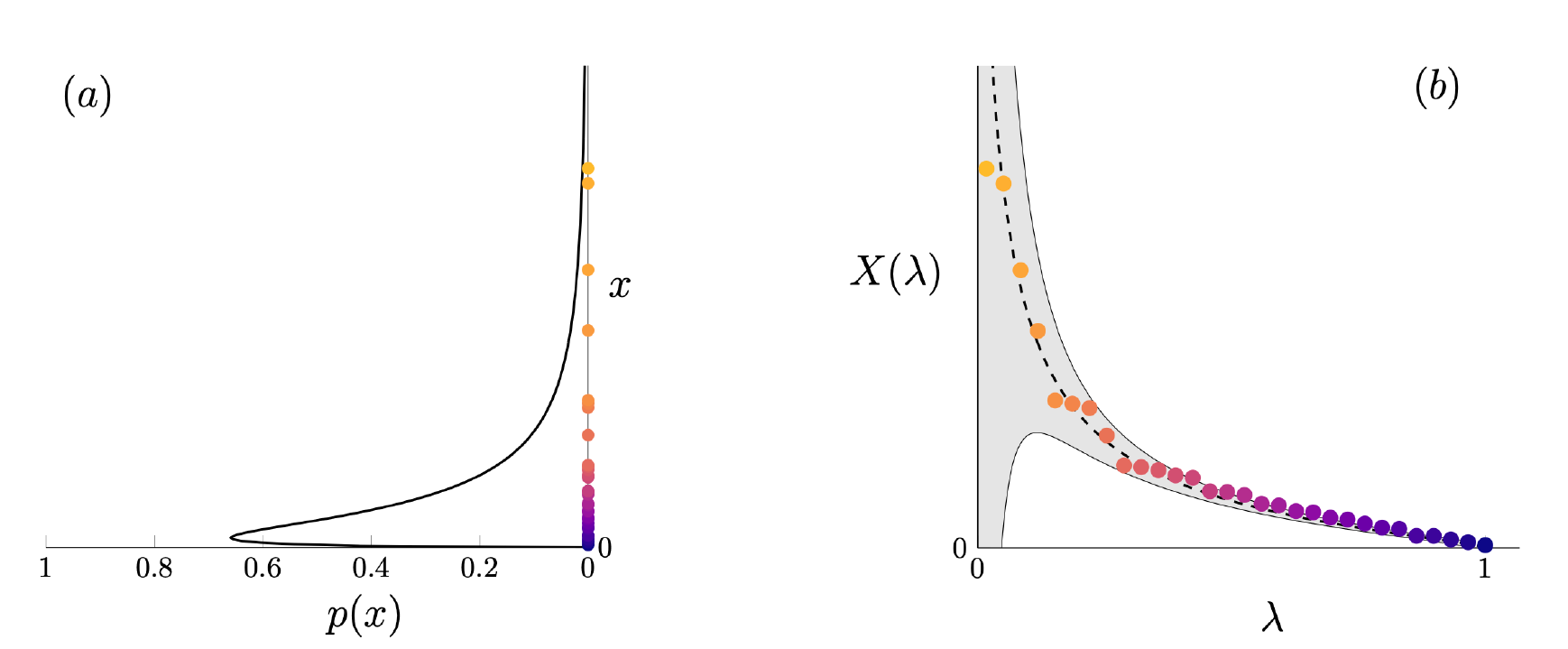}
\caption{Panel (a) represents the parent distribution,
which in this case we arbitrarily chose to be  {\color{black}$p(x)~=~(8/ \pi)\,x^{1/2}\,(x+1)^{-3}$}, 
used to extract $N = 30$ i.i.d. samples, depicted as dots. The same samples are reported in panel (b) {\color{black}after being sorted}, together with the order statistics expectation value $\langle X(\lambda)\rangle $,
obtained from the leading order of Eq.~\eqref{eq:ODR},
depicted as a dashed line. The filled area in panel (b) represents the region within the expected standard deviation $\Sigma^{1/2}\left(X(\lambda),\,X(\lambda) \right)$ from the dashed line,
that was also obtained from Eq.~\eqref{eq:ODR2} at the leading order. }
\label{fig: ODR}
\end{figure*}

\maketitle

\textit{Introduction. -- } In many natural and social phenomena,
items that are ranked according to a certain feature $x$, e.g. their magnitude or frequency, are found to follow an empirical power law $x(r)\sim A/(B+r)^{\alpha}$ (with $r$ being the rank). This is known as {\color{black}Zipf-Mandelbrot law (ZM)} (or, equivalently, Pareto distribution)~\cite{newman2005power,saichev2009theory,clauset2009power}, and is observed rather ubiquitously, e.g. in {\color{black}energies} of earthquakes~\cite{sornette1996rank}, cosmic structures~\cite{de2021zipf}, cities populations~\cite{gabaix1999zipf}, financial returns ~\cite{gabaix2009power}, words usage in a language~\cite{piantadosi2014zipf}, choices of chess openings~\cite{Blasius2009}, visualizations of web pages~\cite{adamic1999nature}, neural activity~\cite{mora2011biological,Tyrcha2013}, the number of citations of scientific papers~\cite{price1965networks} and genes expression~\cite{furusawa2003zipf}, 
to cite but a few.\\
\hspace*{0.25cm} This universality has spurred an intense research aimed at understanding the underlying mechanisms that could lead to the emergence of {\color{black}ZM} law in the organization of such extremely diverse complex phenomena.
Apart from the above-mentioned scenarios where specific models were formulated to explain the observed occurrence on a case-by-case basis, there have recently been efforts to derive {\color{black}ZM} law from less fine-tuned conditions, including hidden variable models ~\cite{Aitchison2016}, stochastic processes~\cite{James2018}, dynamical constraints~\cite{de2021dynamical}, least-effort principles~\cite{Wang2020Principle}, and criticality in multivariate data~\cite{Schwab2014,marsili2022quantifying}.
Despite all these advancements, a single general derivation that would account for the ubiquitous emergence of {\color{black}ZM} law is still missing.
This is precisely what we provide in this work. 
Our result is achieved by making use of notions from the \textit{order statistics}, a field of fundamental importance in understanding the distributional properties of sorted samples~\cite{renyi1953theory,Arnold2008,david2004order}.

First of all, we derive a new general and analytic expression for the cumulants of the order statistics obtained from any {\color{black}large data samples} drawn from a given probability distribution.
Building on this result, 
we then demonstrate that,
whenever a small subset of the ranked samples is considered,
the corresponding order statistics becomes statistically indistinguishable from a ZM distribution.

The consequence of our result is then a rigorous and quantitative proof of the observed ubiquity of ZM law emergence in Nature.

We finally showcase our findings on three relevant examples: (a) the Miller's typing monkey~\cite{miller1957some} (b) the Barab\'{a}si-Albert model~\cite{barabasi1999emergence} and (c) a data sample generated by a Gaussian distribution. In the first two well-known models, we reproduce and extend previously known results as particular applications of our general findings; concerning example (c), we show that normally generated samples can lead to a ZM distribution whenever only a small fraction of the whole sorted dataset is considered.

\textit{Order Duality Relationships. --} 
We start by considering $N$ independently and identically distributed (i.i.d.) draws from {\color{black}an analytic} 
probability density function $p(x)$, henceforth called \textit{parent p.d.f.}, and rank them in non-increasing order,
i.e., $X_{r} \geq X_{r+1}$, $\forall \ r < N$. The resulting sorted set is usually referred to as the `order statistics'~\cite{david2004order}. 
A first crucial observation to make is that the sample size $N$ can be very different and considering a given $X_{r}$ can have a completely different meaning depending on such size. For example, given a sample size $N=9$, $X_5$ would correspond to the median value, while for a larger sample size $N=100$ it would belong to the top $5\%$ outcomes, thus far above the median. 
In order to circumvent this, we introduce a new variable $\lambda = r/N$ which represents the \textit{relative ranking}, and denote $X(\lambda) \equiv X_{N\cdot \lambda}$.
In the limit of $N\to\infty$, $\lambda$ becomes a continuous variable in the interval $ \left(0,1\right]$.

Our first main result is a closed analytic correspondence between all the statistical cumulants of $X(\lambda)$ and the parent p.d.f. (see Supplementary Material, Section A). 
{\color{black}We call these results `Order duality relationships' (ODRs)
and hereby} report the expression for the average 
\begin{equation}\label{eq:ODR}
     \langle X(\lambda)\rangle \!=  F^{-1}\left[1-\frac{N}{N+1} \lambda \right]+ \mathcal{O}\left[\frac{1}{N}\right],
\end{equation}
and for the covariance between order statistics
\begin{align}\label{eq:ODR2}
     \Sigma\left(X\left(\lambda_1\right), X\left(\lambda_2\right)\right) &= \frac{1}{p\left(\langle X\left( \lambda_1 \right) \rangle\right)p\left(\langle X\left( \lambda_2 \right) \rangle\right)}\times\\
     &\,\times\frac{N \lambda_1\left( N[1-\lambda_2]+1\right)}{\left( N+2\right)\left( N+1\right)^2} + \mathcal{O}\left[ \frac{1}{N^2}\right]\,,\notag
\end{align}
where we assumed $\lambda_1 \leq\lambda_2$
In the previous equations we introduced the function $F^{-1}$,
that is the inverse of the cumulative distribution $F(x) = \int_{-\infty}^x dy\, p(y)$.
It is important to stress that the large-$N$ regime we investigate is exactly the framework considered in the works of Zipf ~\cite{zipf2016human}, Mandelbrot~\cite{mandelbrot1953informational} and others~\cite{auerbach2023law, estoup1912gammes, rybski2023auerbach}.
In the $N\to +\infty$ limit, the average value of the order statistics in terms of the relative ranking $\lambda$ tends to a constant,
while their covariance decreases as $\mathcal{O} \left(1/N\right)$.
This has an important consequence,
that is the order statistics exhibit a concentration of measure phenomenon. 
To see this, 
we collect the order statistics,
which are not independent random variables,
in a single random vector $\underline{X} = \left(X\left(\frac{1}{N}\right),\, X\left(\frac{2}{N}\right),\,... \right)$ 
in an $N$-dimensional space, so that we account for their joint behavior. 
We then consider the total probability of the outcomes contained in an $N$-dimensional ball  
centered at $\langle \underline{X} \rangle$
and with a radius $r$.
As $N$ increases, 
we reduce such a radius at the same rate the statistical fluctuations are decreasing,
i.e. $r \sim \mathcal{O}\left( \Sigma^{1/2} \right) \sim \mathcal{O}\left( N^{-1/2} \right)$.
For large $N$, the probability density within this ball behaves as
\begin{equation}
\frac{\mathbb{P}\left[ \underline{X} \in \mathcal{B}_r\left( \langle  \underline{X} \rangle \right) \right]}{\mathrm{Vol}\left( \mathcal{B}_r\left( \langle  \underline{X} \rangle \right) \right)} 
\sim \frac{1}{(N^{-1/2})^{N}} = \mathcal{O}(N^{N/2})\,,
\end{equation}
indicating that the probability is concentrating around $\langle  \underline{X} \rangle$ at a super-exponential rate in $N$.
In Fig.~\ref{fig: ODR} we provide an intuitive illustration of the Order Duality Relationship application.

An immediate effect of the ODRs is that, if the parent p.d.f. is of the form $p(x) \propto x^{-\beta}$ ($\beta \neq 1$), the resulting order statistics expectation value follows a ZM law with $\alpha = (\beta-1)^{-1}$,
in agreement with the result obtained in \cite{cristelli2012there}.
For $\beta = 1$, instead, the order statistics manifests an exponential falloff.
These considerations already provide new insights on several benchmarking models, e.g. the Barab\'{a}si-Albert algorithm~\cite{barabasi1999emergence}, as we will detail in the following Section.

\textit{{\color{black}Local Zipf-Mandelbrot emergence. --}}
Notwithstanding, Eq.~\eqref{eq:ODR} alone still does not account for the apparent ubiquity of {\color{black}ZM} law. 
A first analysis of empirical evidence suggests that such behavior is expected to emerge \textit{locally} when the very top fraction of relative ranking events are considered, which correspond to $\lambda \ll 1$ within our framework. 
For example, in the case of earthquakes
one typically focuses on the detection of very intense magnitude events while neglecting the huge amount of tiny ground vibrations (which could hardly be systematically detected anyway).
Our second finding is to rigorously prove 
that, whenever a small fraction of the order statistics is considered,
the latter becomes statistically indistinguishable from a ZM law.
To this end, we  observe that in the vicinity of any fixed $\lambda_0 = r_0/N$, i.e. for $\lambda \approx \lambda_0$,
it is possible to locally approximate the order statistics with a ZM law.
Concretely, it is possible to find a value for the exponent
$\alpha(\lambda_0) = -1/(z^{(1)}(H_0)+1)$, where $H_0 \equiv \mathrm{ln}(\langle X(\lambda_0)\rangle)$ and 
$z^{(1)}$ is the first derivative of
$z(\eta) \equiv \mathrm{ln}\left( p(e^\eta)\right)$, 
such that 
\begin{equation}\label{eq: local behaviour}
   \langle X(\lambda)\rangle = \frac{A}{\left(B+\lambda\right)^{\alpha}} + \mathcal{O}\left[(\lambda-\lambda_0)^3\right]\,.  
\end{equation}
$A$ and $B$ in the previous equations are constants (we provide their exact but rather cumbersome expressions in the Supplemental Material, Section B).
A crucial consideration stemming from Eq.~\eqref{eq: local behaviour} is that the error associated to such local approximation is third order in the relative ranking $\mathcal{O}\left(\abs{r-r_0}/{N}\right)^3$.
The latter should to be compared with 
the magnitude of the statistical error $\mathcal{O}\left(\Sigma^{1/2} \right)$ coming from the sampling and ordering processes, which we previously discussed to be $\mathcal{O}\left({N^{-1/2}}\right)$ from Eq.~\eqref{eq:ODR}.
Hence, provided that 
\begin{equation}\label{eq: N^5/6}
    \frac{\abs{r-r_0}^3}{N^3} \ll \frac{1}{N^{1/2}} \quad \implies \quad \abs{r-r_0} \ll N^{5/6} \, ,
\end{equation}
the difference between the expected order statistics 
and the ZM law that approximates it
is negligible with respect to statistical fluctuations.
This result implies that the ranking obtained by sorting samples generated with an analytic parent p.d.f. 
becomes statistically indistinguishable from a ZM distribution
whenever a small subset of the whole ranking is considered.
To avoid this, a commonly accepted rule of thumb is that a probability density function should not be considered a power law unless
it exhibits power-law-like behavior over at least a few decades in a feature-rank plot.
Eq.~\eqref{eq: N^5/6} offers a more rigorous criterion,
that is a probability density function should not be \textit{globally} considered a power-law unless
a subset of minimal size $\mathcal{O}\left({N^{5/6}}\right)$ has been considered.
In those situations where this is impractical, 
as in the earthquakes example,
the classification of the order statistics as a global power-law is just infeasible.
As a final comment, it is worth mentioning that the generality of ODRs may seem in contrast with the occasional empirically observed deviations from ZM law at $\lambda \approx 0$ ~\cite{cristelli2012there}.
However, Eq.~\eqref{eq:ODR} do not only agree but provide a recovery of {\color{black}ZM} law also in this regime. The key observation is that the variance of $X(\lambda\approx 0)$ in these instances can diverge as $\lambda/p^2(\langle X(\lambda)\rangle )$ for $\lambda \to 0$. Therefore such deviations become compatible with the ideal expected {\color{black}ZM} behavior within the predicted statistical error.

{\color{black}\paragraph{Discussion. --} It is important to comment on our two main results. First of all, our work does not show that any order statistics is a power-law but that any dataset, once ordered in terms of relative ranking, is described by Eq.~\eqref{eq:ODR}. 
However, in the overwhelming majority of applications, the number of data-points considered of interest is only a small fraction of the total size of the whole sample. For example, when ranking the most populated cities, one typically focuses on the first few hundreds out of tens of thousands of much smaller cities and villages that are not even considered (see e.g. \cite{auerbach2023law}); likewise, when ranking the most energetic earthquakes, one typically completely neglects to even account in the ranking all those very weak events that happen everyday and are even hard to record and concentrates on only the most powerful ones.  
According to the relative ranking variable introduced above, virtually every process considered in the context of Zipf-Mandelbrot-Pareto therefore corresponds to a local neighbourhood $\lambda \approx \lambda_0$. 
In such an interval the expectation value 
of the order statistics can be approximated with a proper ZM distribution as in Eq.~\eqref{eq: local behaviour}.
Since the error of such an approximation becomes negligible 
if compared with the statistical fluctuations predicted by Eq.~\eqref{eq:ODR},
it is practically infeasible to establish if a ranking subset follows a ZM law or a more generic distribution.
}

\textit{Applications. -- }
In the following, we showcase our results on three paradigmatic examples, chosen in light of their interdisciplinarity and wide applicability in several contexts. The first two, in particular, highlight the close relationship between the tail exponents of a power-law parent p.d.f. and the corresponding {\color{black}ZM} law. The third example, instead, shows the aforementioned emergence of local Zipf behaviour for data drawn from a Gaussian parent p.d.f. .

\begin{figure*}[t]
\subfigure[ ]{   \includegraphics[width=0.45\textwidth]{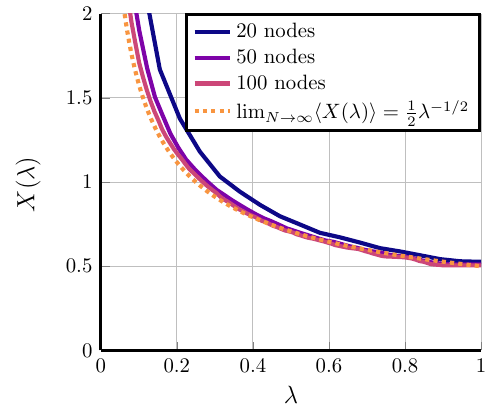}
\label{case_a}}
\subfigure[]{  \includegraphics[width=0.45\textwidth]{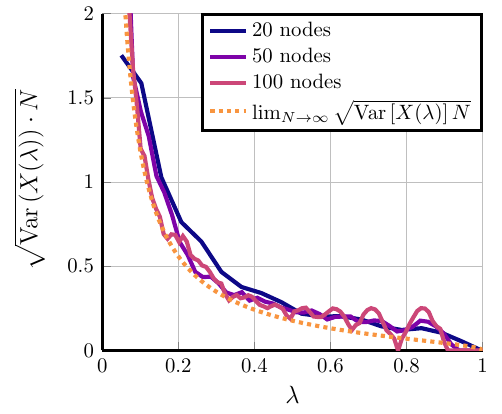}
\label{case_b}}
\caption{{\color{black}Ranking of nodes of a network generated through Barab\'{a}si-Albert algorithm,
where nodes are ranked basing on their number of edges $X$. In the plots $\langle X(\lambda)\rangle$ is normalized on the interval $\lambda \in [0,1]$. 
Panels (a) and (b) represent the expectation values
$\langle X(\lambda)\rangle$ and the standard deviations $\sqrt{\mathrm{Var}\left(X(\lambda)\right)}$ respectively.
The reported results have been obtained numerically for the network sizes $N = \{20, 50, 100 \}$ and 
analytically from Eq.~\eqref{eq:ODR} in the $N \to \infty$ limit. 
The deviations of the numerical results 
from the analytical ones depend on discretization effects (the number of edges is an integer number before normalization), in addition to the finiteness of $N$. Discretization effects are evident in the rightmost part of panel (b).}}
\label{fig: Barabasi-Albert}
\end{figure*}

\textit{1. Miller's Typing monkey.}
We begin by benchmarking our formalism to re-derive one of the classical results associated with {\color{black}ZM} law, known as Miller's typing monkey~\cite{miller1957some}. 
Let $m$ denote the number of letters of a given alphabet and $q_s$ be the probability of hitting the space-bar, thus terminating a word.
Each letter is assumed to have the same probability $(1-q_s)/m$ to be typed
and we remark that any possible string of letters 
will be considered to be a valid word.
In this gedankenexperiment, we imagine a ‘dictionary’ consisting of all possible words,
some of which are typed by the monkey.
One then records the \emph{empirical frequency}, i.e. the relative number of times
each word of the dictionary is actually typed in the finite sequence produced by the monkey.
At the end, one sorts the dictionary words by their empirical frequency
(from most frequent to least frequent), thereby building the order statistics.
Our goal now is to apply our results in order to predict the latter from Eq.~\eqref{eq:ODR}.
It is worth stressing that in this thought experiment the \emph{empirical frequency} $x$
of each word, as observed in the finite sample of \(M\) typed words,
is itself treated as the feature whose order statistics are analyzed.
This makes explicit that our framework naturally covers the case where
frequency plays the role of the feature, thereby connecting to the
traditional frequency–rank formulation of the ZM law.
Thus, the parent p.d.f. \(p(x)\) characterizes how the empirical frequencies
\(x\) of the dictionary words are distributed,
in contrast with the theoretical probabilities \(q(y)\) assigned to words of length \(y\).
We can obtain an expression for $p(x)$ 
by considering a monkey that types \(M\) words. The probability that a single trial produces a particular dictionary word of length \(y\) is
\begin{equation}
q(y)=q_s\!\left(\frac{1-q_s}{m}\right)^y.
\end{equation}
The number of occurrences \(k\) of that word in \(M\) independent trials is binomial; 
for large \(M\) the empirical frequency \(x=k/M\), observed from the finite sample
of \(M\) words, is well approximated by a normal law
with mean \(q(y)\) and variance \(q(y)(1-q(y))/M\).
Now, imagine the frequencies have been recorded in the dictionary,
and we want to obtain the probability of  sampling a word with frequency $x$ from it.
Weighting the previous Gaussian by the number \(m^y\) of words of length \(y\) and changing variables \(y\mapsto q\) yields a Laplace-type integral that is sharply concentrated at \(q=x\) when \(M\gg1\). Applying Laplace's method therefore gives, for \(x\in(0,q_s)\),
\begin{equation}
p(x)\propto x^{\alpha-1}+\mathcal{O}\!\left(\frac{1}{M}\right),\qquad
\alpha=\frac{\ln m}{\ln\!\big(\tfrac{1-q_s}{m}\big)}
\end{equation}
(see Section C in the Supplemental Material for the step-by-step derivation).
Thus, in the large-\(M\) limit the distribution of empirical frequencies across dictionary words
approaches a power law
\(p(x)\propto x^{\alpha-1}\) on \((0,q_s)\), with a controlled total error of order \(1/M\).
Since the parent distribution is a power-law $p(x) \propto x^{-\beta}$
with exponent
\begin{equation}
    \beta = 1+\frac{b}{a},
\end{equation}
its domain has to be restricted to an interval interval $[\epsilon, 1]$,
for a fixed cutoff $\epsilon > 0$,
in order to be normalizable.
At this point we can apply Eq.~\eqref{eq:ODR}
to obtain that expectation value 
and the variance of the associated order statistics
\begin{equation}
    \lim_{N\to \infty}\langle X(\lambda)\rangle = \left[1 +\left(\epsilon^{-b/a} -1\right)\lambda\right]^{-a/b}
\end{equation}
that is a ZM law with an exponent that
that agrees with and extends the result originally obtained by Miller ~\cite{miller1957some}, which only described the average behavior of sets of words with equal length.
Moreover, 
in addition to the original result of Miller,
Eq.~\eqref{eq:ODR2} allows predicting the covariances of the order statistics
\begin{align*}
    \lim_{N\to \infty}N \Sigma (X\left(\lambda_1\right)&, X\left(\lambda_2 \right)) = \frac{\lambda_1}{\left[\epsilon^{-b/a} -\frac{b(1-\lambda_1)}{a}\right]^{\left(a/b+1 \right)}}\times\\
    &\quad\quad\quad\quad \times \frac{(1-\lambda_2)}{\left[\epsilon^{-b/a} -\frac{b(1-\lambda_2)}{a}\right]^{\left(a/b+1 \right)}}\,.
\end{align*}
for $\lambda_1 \leq \lambda_2$.\\

\textit{2. Barab\'{a}si-Albert model}
Several real world networks 
turn out to be scale-free~\cite{caldarelli2007scale,albert2002statistical}.
These include, e.g., the internet~\cite{pastor2001epidemic}, cellular structures~\cite{albert2005scale} and scientific coauthorship networks~\cite{newman2001structure}, among many others. This is equivalent to the statement 
that a randomly picked node of the network 
has a probability
\begin{equation}
  p(x)\propto x^{-\beta}, \quad \beta > 0 
\end{equation}
to have a number of edges $x$. The
Barab\'{a}si-Albert (BA) model provides an algorithm,
based on the concepts of \textit{growth} and \textit{preferential attachment} \cite{barabasi1999emergence,albert2002statistical},
that allows to construct a network with this feature.
It is known that, independently from initial conditions, BA networks exhibit $\beta = 3$ as the number of nodes grows very large.
More complex models have been proposed in order to produce networks with $\beta \neq 3$~\cite{kaprivsky-redner}.
We can now identify $p(x)$ as the parent distribution that assigns to each node its number of edges. 
The associated order statistics 
describes the variation of the edges number
as we run down the ranking of the 'most connected nodes'.
Using the ODRs Eq.~\eqref{eq:ODR} and Eq.~\eqref{eq:ODR2},
we can predict the behavior
$\langle X(\lambda) \rangle \sim \lambda^{-1/(\beta-1)}$
for such an order statistics.
In particular,
for the BA model one obtains $\langle X(\lambda) \rangle \sim \lambda^{-1/2}$ for the expectation value and $\mathrm{Var}\left[ X(\lambda)\right]\sim (\lambda-1)/(N\lambda^2)$
for its variance.
We test our prediction numerically
by using the BA algorithm 
to independently generate $100$ networks 
with a total number of nodes $N$,
which are then sorted 
based on their number of edges.
This allows us to calculate the mean and the first two cumulants of the order statistics.
We report the results for $N = \{20, 50, 100 \}$
in Fig.~\ref{fig: Barabasi-Albert}, 
where we observe the convergence to the analytical results obtained from Eq.~\ref{eq:ODR} 
in the $N \to \infty$ limit, with a standard deviation that exhibits a $\sim 1/\sqrt{N}$ falloff.

\begin{figure}[t]
\includegraphics[width=0.48\textwidth]{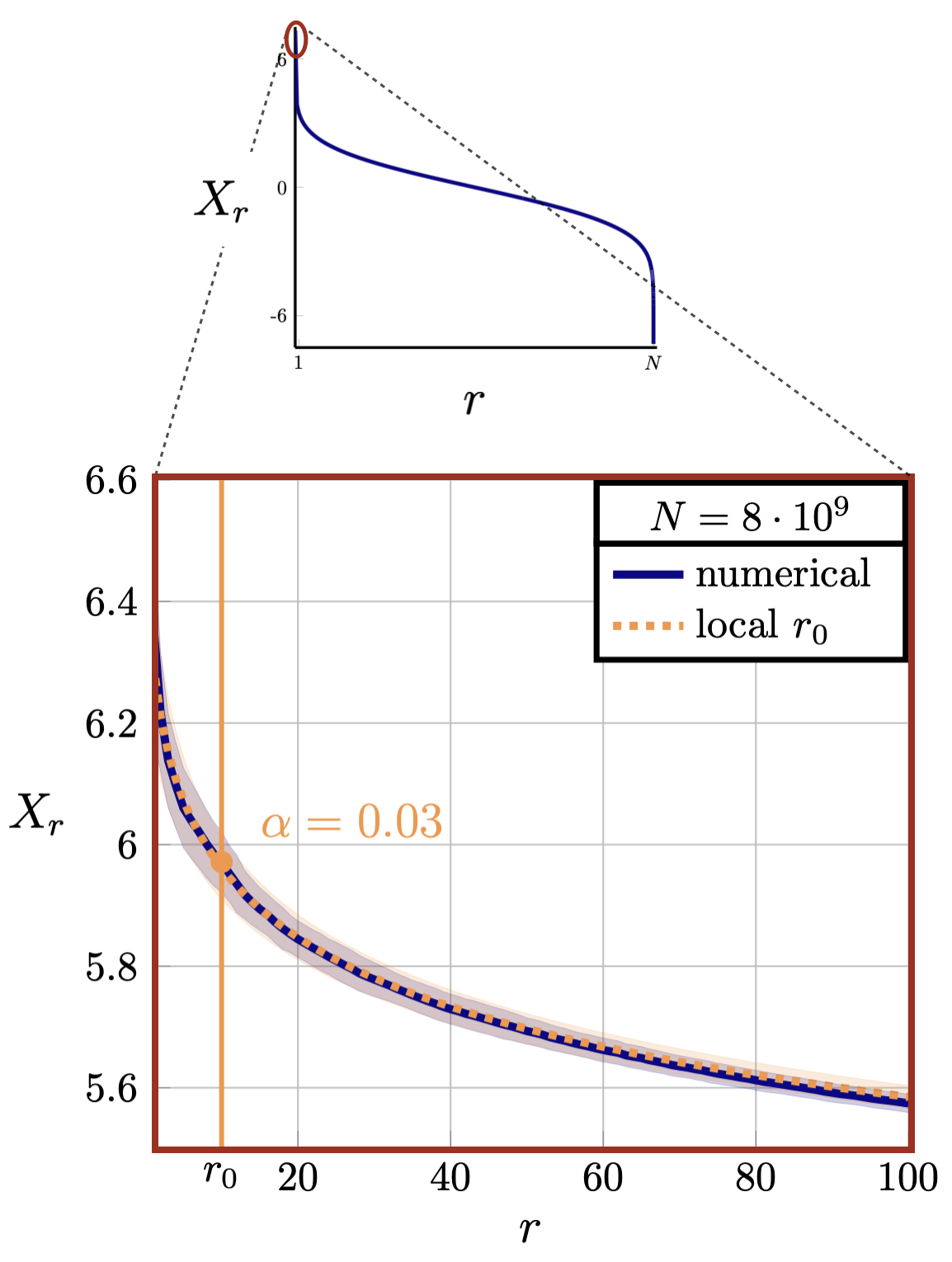}
\caption{{\color{black}Order statistics $X_r$ associated with a Gaussian parent distribution with $\mu = 0$ and $\sigma = 1$.
Numerical results are reported together with shades representing their statistical error.
The plot represents the top $100$ ranked draws out of  $N = 8 \cdot 10^9$ total samples.
The local power-law behaviour around $r_0 = 10$,
theoretically predicted from Eq.~\eqref{eq: local behaviour},
is reported as a dashed line together with its predicted statistical fluctuations obtained from Eq.~\eqref{eq:ODR}.}}
\label{fig:Gaussian}
\end{figure}

\textit{3. Normal distribution.} We now turn our attention to showcasing the notion of local ZM emergence. Given its ubiquity --- due to the Central Limit Theorem --- the order {\color{black}statistics associated to a Gaussian parent p.d.f.
\begin{equation}
    p(x; \mu, \sigma) = \frac{1}{\sqrt{2\pi}\sigma}e^{-(x-\mu)^2/2\sigma^2} 
\end{equation}
is of paramount importance.}
Using the ODRs 
one can directly obtain 
\begin{align*}
    \langle X(\lambda)\rangle \,&\mathop\sim^{N\to \infty} \,\mu +\sqrt{2}\sigma\,\mathrm{erf}^{-1}\left(1-2\lambda\right)\\
    N\Sigma\left(X\left(\lambda_1\right), X\left(\lambda_2\right)\right)  \,&\mathop\sim^{N\to \infty} \,2\pi \sigma^2\lambda_1e^{\left(\mathrm{erf}^{-1}\left(1-2\lambda_1\right)\right)^2}\times\\
    & \times(1-\lambda_2)e^{\left(\mathrm{erf}^{-1}\left(1-2\lambda_2\right)\right)^2},    
\end{align*}
where $\mathrm{erf}^{-1}$ is the inverse of the so-known “error function”.
Moreover, 
it is possible to apply our second result 
to give an analytical local description of the order statistics,
approximating $\langle  X(\lambda) \rangle$ with a ZM distribution around any given $\lambda_0$, as described in Eq.~\eqref{eq: local behaviour}.
In particular, for a normal parent p.d.f. the exponent $\alpha(\lambda_0)$ is given by
\begin{equation}\label{eq: gauusian exponent}
    \alpha(\lambda_0) = \left[ \frac{\langle X(\lambda_0)\rangle \left(\langle X(\lambda_0)\rangle-\mu \right)}{\sigma^2}-1\right]^{-1} \, .
\end{equation}
There are only two isolated values of $\lambda_0$ 
for which Eq.~\eqref{eq: gauusian exponent} is singular.
In these particular cases 
the local behavior is exponential rather than power-like
(see Supplemental Material Section D for details).

Fig.~\ref{fig:Gaussian}, 
illustrates the expected values of the order statistics for the top $100$ samples 
drawn from a dataset of size  $N = 8 \cdot 10^9$, 
which approximates the global population at the time of writing. 
These values are generated from the given Gaussian probability density function (solid line) 
and correspond to the top $0.00001 \%$ of the ranking.
Additionally, 
we present a local approximation of the order statistics, 
derived from Eq.~\eqref{eq:ODR} and Eq.~\eqref{eq: local behaviour}.
This focus on the highest percentiles of a ranking is relevant in many real-world scenarios involving large datasets. For example, when ranking individuals worldwide based on a specific characteristic, attention is typically restricted to the top-performing subset.

\textit{Conclusions. -- } In our work we have obtained a novel analytical correspondence between a generic probability distribution and its associated order statistics. We then made use of this result, which we called Order Duality Relationships (ODRs), {\color{black}to rigorously show that
in the large sample size limit
any analytic distribution cannot be locally discerned from a ZM law.
This provides a demonstration of the universality of {\color{black}ZM} emergence in Nature, which we then proceeded to showcase in three paradigmatic examples.}

The first model we considered is the Miller's typing monkey problem, 
that allowed us to compare our results with another analytical benchmark
that has been reproduced and extended.
{\color{black}The second is the Barab\'{a}si-Albert model, 
for which we predicted the order statistics' first two cumulants and compared them with the corresponding numerically-obtained estimators.}
Finally, 
we moved our attention to the case of Gaussian parent p.d.f.,
that is of extreme interest for most real-word applications.
In this case we further verified that the local power-like approximation 
of the parent p.d.f. leads to a satisfactory description of the order statistics,
whenever a small fraction of the full ranking is taken into account.

\bibliographystyle{apsrev4-1}
\bibliography{main}

\newpage

\appendix

\widetext

\newpage 
\begin{center}
\vskip0.5cm
{\Large Supplemental Material}
\end{center}
\vskip0.4cm

\setcounter{section}{0}
\setcounter{equation}{0}
\setcounter{figure}{0}
\setcounter{table}{0}
\setcounter{page}{1}
\renewcommand{\theequation}{S\arabic{equation}}
\renewcommand{\thefigure}{S\arabic{figure}}

In this Supplemental Material we provide the proofs and all additional details necessary to obtain the results presented in the main text.
In particular, Section A is devoted to the analytical derivation od the Order Duality Relationships, viz. Eq.~\eqref{eq:ODR} of the main text; in Section B we demonstrate the local indistinguishability of a generic order statistics from a Zipf's law, viz. Eq.~\eqref{eq: local behaviour} of the main text; Section C finally contains additional details and considerations about the Zipf's behavior associated to a Gaussian parent p.d.f. .

\section{Section A: Order Duality Relations}\label{sec: ODR}
In this section we prove the Order Duality Relations
in Eq.~\eqref{eq:ODR}.
At this scope, 
we first focus on the simple scenario
of a uniformly distributed parent p.d.f.,
whose results are then extended 
to more generic parent distributions.
Let $g(u)$ be the uniform distribution defined by
\begin{equation}
    g(u) = \begin{cases}
        1,& \mathrm{if} \; u \in [0,1]\\
        0, & \mathrm{otherwise}\\
    \end{cases}
\end{equation}
and let $\{u_r\}_{r \in \{1, 2, ..., N\}}$ be a set of $N$ 
independent random variables distributed according to $g(u)$.
We denote with $\{U_r\}$
the order statistic obtained by sorting the $\{u_r\}$ in non-increasing order.
The p.d.f. for the r-th order statistic $U_r$ is
\begin{equation}
    g_{r}(u) = \frac{N!}{(N-r)!(r-1)!}[1-u]^{r-1} u^{N-r} \,.
\end{equation}
The expectation value for the k-th moment 
can be obtained by straightforward calculation 
and is given by
\begin{equation}\label{eq: U^k}
    \langle U_{r}^k \rangle = \frac{N!}{(N+k)!}\frac{(N+k-r)!}{(N-r)!}\quad\quad \forall k \in \mathds{N}.
\end{equation}
We hereby introduce the change of variable
\begin{equation}
    \lambda \equiv \frac{r}{N} \in \left\{\frac{1}{N}, \frac{2}{N}, .. 1\right\}
\end{equation}
and use the notation $ U(\lambda) \equiv U_{\lambda N}$,
in terms of which
\begin{equation}\label{eq: U moments}
    \langle U^k(\lambda)\rangle= \frac{N!}{(N+k)!}\frac{(N[1-\lambda]+k)!}{(N[1-\lambda])!}\,.
\end{equation}
This change of variable allows us to study the $N\to \infty$ limit
while keeping the ratio $\lambda = r/N$ fixed.
This has as a direct consequences
\begin{equation}\label{eq: expectation value}
    \langle U(\lambda)\rangle= 1- \frac{N}{N+1}\lambda \, ,
\end{equation}
\begin{equation}\label{eq: 2 power}
    \left\langle (U(\lambda)-\langle U(\lambda)\rangle)^2\right\rangle= \frac{\lambda N(N[1-\lambda]+
    1)}{(N+1)^2(N+2)} \, ,
\end{equation}
and 
\begin{equation}\label{eq: p power} 
    \left\langle (U(\lambda)-\langle U(\lambda)\rangle)^p\right\rangle= \mathrm{o}\left( \frac{1}{N}\right) \quad\quad \forall p \in \mathds{N},\, p > 2\, ,
\end{equation}
that will be extensively used in what follows.
We now proceed with the computation of covariance 
for order statistics obtained from uniform sampling,
namely
\begin{equation}
    \Sigma\left(U_{r_1}, U_{r_2}\right) = \langle U_{r_1} U_{r_2} \rangle - \langle U_{r_1}\rangle \langle U_{r_2}\rangle \,,
\end{equation}
where we assume $0 \leq r_1 < r_1 \leq N$.
From Eq.~\eqref{eq: U^k} we already know that
\begin{equation}
    \langle U_{r_1}\rangle \langle U_{r_2}\rangle = \frac{N+1-r_1}{N+1}\frac{N+1-r_2}{N+1}\,.
\end{equation}
We therefore proceed by computing
\begin{equation}
\begin{split}
    \langle U_{r_1} U_{r_2} \rangle
    &= \int_0^1 du_1 \int_0^{u_1}du_2\, g(u_1,u_2)\,u_1 u_2\\
    &= \int_0^1 du_1 \int_0^{u_1}du_2\, \frac{N![1-u_1]^{r_1-1}[u_1-u_2]^{r_2-r_1-1}u_2^{N-r_2}}{(r_1-1)!(r_2-r_1-1)!(N-r_2)!}\,u_1 u_2\\
    &= \frac{N!}{(r_1-1)!(r_2-r_1-1)!(N-r_2)!}\int_0^1 du_1 \, u_1[1-u_1]^{r_1-1}\int_0^{u_1}du_2\, [u_1-u_2]^{r_2-r_1-1}u_2^{N+1-r_2}\\
    &= \frac{N!}{(r_1-1)!(r_2-r_1-1)!(N-r_2)!}\int_0^1 du_1 \, u_1^{N+2-r_1}[1-u_1]^{r_1-1}\int_0^{1}dz\, [1-z]^{r_2-r_1-1}z^{N+1-r_2}\\
    &= \frac{N!}{(r_1-1)!(r_2-r_1-1)!(N-r_2)!}\beta\left(N+3-r_1, r_1 \right)\beta\left( r_2-r_1, N+2-r_2\right)\\
    &= \frac{N!}{(r_1-1)!(r_2-r_1-1)!(N-r_2)!}\frac{\Gamma\left( N+3-r_1\right)\Gamma\left( r_1\right)}{\Gamma\left( N+3\right)}
    \frac{\Gamma\left( r_2-r_1\right)\Gamma\left( N+2-r_2\right)}{\Gamma\left( N+2-r_1\right)}\\
    &= \frac{N!}{(r_1-1)!(r_2-r_1-1)!(N-r_2)!}\frac{\left( N+2-r_1\right)!\left( r_1-1\right)!}{\left( N+2\right)!}
    \frac{\left( r_2-r_1-1\right)!\left( N+1-r_2\right)!}{\left( N+1-r_1\right)!}\\
    &= \frac{\left( N+2-r_1\right)\left( N+1-r_2\right)}{\left( N+2\right)\left( N+1\right)}\,,
\end{split}
\end{equation}
that allows to obtain the expression for the covariance
\begin{equation}
    \Sigma\left(U_{r_1}, U_{r_2}\right) = \frac{r_1\left( N+1-r_2\right)}{\left( N+2\right)\left( N+1\right)^2}\,.
\end{equation}
Moving to the relative variables $\lambda_i = r_i/N$, the previous equation reads
\begin{equation}\label{eq: uniform cov}
    \Sigma\left(U\left(\lambda_1\right), U\left(\lambda_2\right)\right) = \frac{N\lambda_1\left( N[1-\lambda_2]+1\right)}{\left( N+2\right)\left( N+1\right)^2}\,,
\end{equation}
that clearly is $\mathcal{O}\left(\frac{1}{N} \right)$, as stated in the main text.\\

We now move our attention to the more generic framework 
of independent random variables $\{x_r\}_{r\in\{1,...,N\}}$
whose associated analytic p.d.f. is $p(x)$.
Let $F(x) = \int_{-\infty}^x dx'\,p(x')$ be the correspondent cumulative density function (c.d.f.)
that can be inverted since $p(x)$ is analytic.
Then, the random variables $u_r \equiv F(x_r)$ 
are uniformly distributed between 0 and 1
and therefore the results obtained above hold.
The independent variables  
\begin{equation}\label{eq: X definition}
    X_{r} \equiv F^{-1}\left( U_{r}\right)
\end{equation}
obtained from the order statistics $\{U_{r}\}$ are themselves 
order statistics, i.e.
\begin{equation}
    X_{r} \geq X_{r+1}.
\end{equation}
Using the same notation introduced for the uniform parent scenario,
we write $X(\lambda) = X_{\lambda N}$.
Starting from eq.~\eqref{eq: X definition}
and Taylor-expanding around $\langle U(\lambda)\rangle $
we obtain
\begin{equation}
      \langle X^k(\lambda)\rangle=  \left\langle\left[ F^{-1}\left(U(\lambda)\right)\right]^k\right\rangle 
       =  \left\langle \left(\Sigma_{j=0}^\infty \frac{1}{j!}\frac{d^jF^{-1}}{dU^j}\Big|_{\langle U(\lambda)\rangle}\cdot \left(U(\lambda)- \langle U(\lambda)\rangle\right)^j\right)^k\right\rangle
\end{equation}
We can know use eq.~\eqref{eq: p power} and eq.~\eqref{eq: 2 power}
to obtain 
\begin{equation}\label{eq: E(X)}
    \boxed{\langle X(\lambda)\rangle =  F^{-1}\left[\langle U(\lambda)\rangle\right]+ \mathrm{o}\left( 1\right)}\, ,
\end{equation}
that is the first equation of the Order Duality Relationships.
We henceforth move to the computation of the covariance
\begin{equation}
    \Sigma\left(X_{r_1}, X_{r_2}\right) = \langle X_{r_1} X_{r_2} \rangle - \langle X_{r_1}\rangle \langle X_{r_2}\rangle\,.
\end{equation}
Leveraging again the fact that $X_r = F^{-1}\left( U_r\right)$
for uniform order statistics $\left\{U_r\right\}_{r=1}^N$,
we can Taylor expand each term as
\begin{equation}
\begin{split}
    \Sigma\left(X_{r_1}, X_{r_2}\right)
    &= \langle F^{-1}\left(U_{r_1}\right) F^{-1}\left(U_{r_2}\right) \rangle - \langle F^{-1}\left(U_{r_1}\right)\rangle \langle F^{-1}\left(U_{r_2}\right)\rangle\\
    &= \sum_{l=0}^\infty \sum_{m=0}^\infty \frac{1}{l!m!}\frac{d^lF^{-1}}{dU^l}\Big|_{\langle U_{r_1}\rangle }\frac{d^mF^{-1}}{dU^m}\Big|_{\langle U_{r_2}\rangle }\times \\
    & \quad\quad \times\left[\langle\left(U_{r_1}-\langle U_{r_1}\rangle\right)^l \left(U_{r_2}-\langle U_{r_2}\rangle\right)^m \rangle - \langle\left(U_{r_1}-\langle U_{r_1}\rangle\right)^l\rangle \langle \left(U_{r_2}-\langle U_{r_2}\rangle\right)^m \rangle \right]\,.
\end{split}
\end{equation}
Since we are interested in the asymptotic behavior,
we focus on the last factor of the sum,
i.e.
\begin{equation} 
    \langle\left(U_{r_1}-\langle U_{r_1}\rangle\right)^l \left(U_{r_2}-\langle U_{r_2}\rangle\right)^m \rangle - \langle\left(U_{r_1}-\langle U_{r_1}\rangle\right)^l\rangle \langle \left(U_{r_2}-\langle U_{r_2}\rangle\right)^m \rangle \,.
\end{equation}
We observe that if $l = 0$ or $m  = 0$ the two terms become identical and the result is 0.
If $l=1$ or $m = 1$ the second term vanishes, 
in particular if $l = m = 1$ the expression reduces to the covariance of Eq.~\eqref{eq: uniform cov}
which we proved to be $\mathcal{O}\left( \frac{1}{N}\right)$ once switched to the relative ranking variable.
Finally, if $l > 1$ or $m >1$, using the Cauchy-Schwarz inequality
\begin{equation}
\begin{split}
     \langle\left(U_{r_1}-\langle U_{r_1}\rangle\right)^l \left(U_{r_2}-\langle U_{r_2}\rangle\right)^m \rangle 
     &\leq \left[ \langle\left(U_{r_1}-\langle U_{r_1}\rangle\right)^{2l}\rangle \langle \left(U_{r_2}-\langle U_{r_2}\rangle\right)^{2m} \rangle \right]^{1/2}\\
     &= \left[ \mathrm{o}\left(\frac{1}{N^2} \right) \right]^{1/2}\\
     &= \mathrm{o}\left(\frac{1}{N} \right)\,.
\end{split}
\end{equation}
As a consequence
\begin{equation}
\begin{split}
    \Sigma\left(X\left( \lambda_1 \right), X\left( \lambda_2 \right)\right) 
    &=  \frac{dF^{-1}}{dU}\Big|_{\langle F\left(X\left( \lambda_1 \right)\right)\rangle}\frac{dF^{-1}}{dU}\Big|_{\langle F\left(X\left( \lambda_2 \right)\right)\rangle }\mathrm{Cov}\left(F\left(X\left( \lambda_1 \right)\right), F\left(X\left( \lambda_2 \right)\right)\right) + \mathrm{o}\left(\frac{1}{N} \right) \\
    &=  \frac{1}{p\left(\langle X\left( \lambda_1 \right) \rangle\right)p\left(\langle X\left( \lambda_2 \right) \rangle\right)}\frac{N \lambda_1\left( N[1-\lambda_2]+1\right)}{\left( N+2\right)\left( N+1\right)^2} + \mathrm{o}\left(\frac{1}{N} \right)\,,
\end{split}
\end{equation}
proving the second order duality relationship.
Remarkably,
$\Sigma\left(X\left( \lambda_1 \right), X\left( \lambda_2 \right)\right)  \sim \mathcal{O}\left( \frac{1}{N}\right)$
independently from the distribution $p(x)$.

\clearpage

\section{Section B: Local Zipf behaviour}
~\label{sec: Local}
We hereby start from Eq.~\eqref{eq: E(X)} 
to show that any order statistics can be locally described 
by a Zipf Law.
In particular, once fixed $\lambda_0\in [0,1]$, one gets
\begin{equation}
    \frac{N}{N+1}\left(\lambda_0-\lambda\right) = \int_{\langle X(\lambda_0)\rangle}^{\langle X(\lambda)\rangle}p(x)dx.
\end{equation}
In what follows we assume that both $\langle X(\lambda)\rangle$ and $\langle X(\lambda_0)\rangle$ are positive, 
analogous results can be obtained for negative values
after the redefinition $x \mapsto -x$.
We consider the change of variables 
\begin{equation}\label{eq: change of variables}
\begin{cases}
    x = e^\eta\\
    p(x) = e^{z(\eta)}
\end{cases}
\end{equation}
From the the assumption of $p(x)$ to be analytic
it follows that also $z(\eta)$ is analytic.
As a consequence $z(\eta)$ is smooth and  
its Taylor expansion around any point converges in an interval.
Notice that the parent distribution $p(x)$ 
is a power-law iff the derivatives $z^{(i)}(H_0)$ vanish $\forall i\geq 2$.
Indeed, let $p(x) \propto x^{-\beta}$, then
\begin{equation}
\frac{p(x)}{p(x_1)} = \left(\frac{x}{x_1}\right)^{-\beta}
\end{equation}
After the change of variables defined in  Eq.~\eqref{eq: change of variables},
the previous equation becomes
\begin{equation}
    z(\eta) = z(\eta_1) - \beta (\eta-\eta_1), \quad \quad x_1 \equiv e^{\eta_1},
\end{equation}
that is a linear behaviour and therefore $z^{(i)}(H_0) = 0, \;\forall i\geq 2$.
We now define
$H(\lambda) = \mathrm{ln}(\langle X(\lambda)\rangle)$
and
$H_0 = \mathrm{ln}(\langle X(\lambda_0)\rangle)$,
obtaining
\begin{equation}\label{eq: all orders equation}
\begin{split}
    \frac{N}{N+1}\left(\lambda_0-\lambda\right) &= \int_{H_0}^{H(\lambda)}d\eta \,\mathrm{exp}\left[z(\eta) + \eta \right] =\\
    &= \mathrm{exp}\left[z(H_0) + H_0 \right]\int_{H_0}^{H(\lambda)}d\eta\,\mathrm{exp}\left[(z^{(1)}(H_0) + 1)(\eta - H_0)+ \sum_{i = 2}^\infty \frac{z^{(i)}(H_0)}{i!} \, [\eta - H_0]^i\right].
\end{split}
\end{equation}
The idea is now to  Taylor-expand both sides around $\lambda = \lambda_0$ 
and compare them term by term.
From the first order we obtain
\begin{equation}
    \frac{dH}{d\lambda}\Big|_{\lambda_0}  = -\frac{N}{N+1}\mathrm{exp}\left[-\left(z(H_0) + H_0 \right)\right]\,.\,
\end{equation}
Moving to the second order we get
\begin{equation}
    \frac{d^2H}{d\lambda^2}\Big|_{\lambda_0} = -(z^{(1)}(H_0) + 1)\left( \frac{N}{N+1}\right)^2\mathrm{exp}\left[-2\left(z(H_0) + H_0 \right)\right]\,.\,
\end{equation}
Finally, we repeat the calculation for the third order, obtaining
\begin{equation}
    \frac{d^3H}{d\lambda^3}\Big|_{\lambda_0} = -\left[2(z^{(1)}(H_0) + 1)^2 - z^{(2)}(H_0) \right]\left( \frac{N}{N+1}\right)^3\mathrm{exp}\left[-3\left(z(H_0) + H_0 \right)\right]\,.\,
\end{equation}
Observe that if we approximate $p(x)$ 
with a power law around $\langle X(\lambda_0) \rangle$,
i.e. we set $z^{(i)}(H_0) = 0$ for all $i\geq 2$,
the result obtained for $H(\lambda)$
from Eq.~\eqref{eq: all orders equation}
doesn't change, up to third order corrections.
Since $\langle X(\lambda)\rangle = \mathrm{exp}\left[H(\lambda)\right]$, 
the same property holds for $\langle X(\lambda)\rangle$.
Therefore, 
one can approximate the p.d.f. as
\begin{equation}
\begin{split}
    p(x) &\approx \mathrm{exp}\left[z\left(H_0\right)+ z^{(1)}\left(H_0\right) \left(\eta-H_0\right) \right] \\
    &= p\left( \langle X(\lambda_0)\rangle\right)\left[\frac{x}{\langle X(\lambda_0)\rangle}\right]^{z^{(1)}\left(H_0\right)}
\end{split}
\end{equation}
and use it to solve Eq.~\eqref{eq: all orders equation},
obtaining an expression for $\langle X(\lambda)\rangle$
that is known to be correct up to third order terms.
Explicitly, one obtains
\begin{equation}\label{eq: local behaviour explicit}
\langle X(\lambda)\rangle = \left[ \langle X(\lambda_0)\rangle^{z^{(1)}(H_0)+1} - (z^{(1)}(H_0)+1)(\lambda-\lambda_0)\right]^{1/(z^{(1)}(H_0)+1)}+ \mathcal{O}\left[\left( \lambda-\lambda_0\right)^3\right]
\end{equation}
that can be cast as in Eq.~\eqref{eq: local behaviour} $ \langle X(\lambda)\rangle = A/\left(B+\lambda\right)^{\alpha} + \mathcal{O}\left[(\lambda-\lambda_0)^3\right]\,$
upon the following identifications:
\begin{align}\label{eq: local alpha}
    \alpha &\equiv \alpha(\lambda_0) = -1/(z^{(1)}(H_0)+1) \, ,\\
    A &\equiv \alpha^\alpha \, ,\\
    B &\equiv \alpha \langle X(\lambda_0)\rangle^{-1/\alpha}-\lambda_0 \,.
\end{align}

\clearpage

\section{Section C: Miller's typing monkey}
~\label{sec: Miller}
\noindent
In the main text we benchmarked our formalism against the classical
Miller's typing monkey experiment~\cite{miller1957some}.
In this thought experiment, a monkey types words at random
from an alphabet of size \(m\), with probability \(q_s\) of hitting the space bar
to terminate a word.
Each of the \(m\) letters is selected independently with probability
\((1-q_s)/m\),
so that any possible string of letters constitutes a valid dictionary word.
Recording the frequencies with which dictionary words are typed,
one can then order them from most to least frequent and compare
to the predictions of Zipf--Mandelbrot laws.
Our goal in this section is to provide the detailed derivation of the parent
frequency distribution \(p(x)\) that underlies this experiment,
justifying the result summarized in the main text.
Consider a monkey that types \(M\) words. The probability that the outcome of a trial is a particular word \(w\) of length \(y\) is
\begin{equation}\label{eq:q_of_y}
    q(y) \equiv q_s\left(\frac{1-q_s}{m}\right)^y .
\end{equation}
Accordingly, the probability that \(w\) occurs exactly \(k\) times over the \(M\) independent trials is the binomial
\begin{equation}
    \mathcal{B}(k;M,q(y))=\binom{M}{k}q(y)^k\big(1-q(y)\big)^{M-k}.
\end{equation}
By the Central Limit Theorem, for large \(M\) the binomial distribution is well approximated by a Gaussian:
\begin{equation}
    \mathcal{B}(k;M,q(y))
    =\mathcal{N}\!\big(k;Mq(y),\,Mq(y)(1-q(y))\big)
    +\mathcal{O}\!\left(\frac{1}{M}\right),
\end{equation}
with mean \(Mq(y)\) and variance \(Mq(y)(1-q(y))\).
Introduce the empirical frequency
\[
    x=\frac{k}{M}.
\]
In terms of \(x\) the Gaussian approximation becomes
\begin{equation}\label{eq:P_x_given_q}
    P(x;\,q(y))=\mathcal{N}\!\left(x;\,q(y),\,\frac{q(y)(1-q(y))}{M}\right)
    +\mathcal{O}\!\left(\frac{1}{M}\right),
\end{equation}
i.e. a normal density with mean \(q(y)\) and variance \(q(y)(1-q(y))/M\).
Now, imagine all the empirical frequencies have been recorded in a dictionary.
We seek the probability density \(p(x)\) for the recorded empirical frequency \(x\) when picking uniformly at random a dictionary word. 
Weighting by the number of words of length \(y\) (which grows like \(m^y\)) and integrating over \(y\) gives
\begin{align}\label{eq:px_integral_y}
    p(x) &\propto \int dy \; m^y \, P(x;\,q(y)) \notag\\
         &= \int dy \; m^y \, \mathcal{N}\!\left(x;\,q(y),\,\frac{q(y)(1-q(y))}{M}\right)
            +\mathcal{O}\!\left(\frac{1}{M}\right)\,. 
\end{align}
Change integration variables from \(y\) to \(q\) using \eqref{eq:q_of_y}. Define
\[
a\equiv\frac{1-q_s}{m}\quad(0<a<1),\qquad q=q_s a^y.
\]
Then
\[
y=\frac{\ln(q/q_s)}{\ln a},\qquad m^y=\left(\frac{q}{q_s}\right)^{\frac{\ln m}{\ln a}},\qquad
dy=\frac{dq}{q\ln a}.
\]
Absorbing constant factors into the proportionality, \eqref{eq:px_integral_y} can be written as
\begin{equation}\label{eq:px_integral_q}
    p(x)\propto \int_{0}^{q_s} dq \; q^{\alpha-1}\,
    \sqrt{\frac{M}{2\pi q(1-q)}}\,
    \exp\!\left[-\frac{M(x-q)^2}{2q(1-q)}\right]
    +\mathcal{O}\!\left(\frac{1}{M}\right),
\end{equation}
where we have introduced the exponent
\[
\alpha \equiv \frac{\ln m}{\ln a}=\frac{\ln m}{\ln\!\big(\tfrac{1-q_s}{m}\big)}.
\]
To perform an asymptotic evaluation, it is convenient to extend the integral to the whole real line. Assume \(x\in(0,q_s)\). The contribution from the tails
\[
\varepsilon_{\pm}=\int_{\pm A}^{\pm\infty}\!\!dq\; 
|q|^{\alpha-1}\sqrt{\frac{M}{2\pi q(1-q)}}
\exp\!\left[-\frac{M(x-q)^2}{2q(1-q)}\right],
\]
for some \(A\in(0,q_s)\), can be controlled by Cauchy--Schwarz:
\begin{align*}
\varepsilon_{\pm}
&\le \left[\int_{\pm A}^{\pm\infty}\!\!dq\;|q|^{2\alpha-2}\sqrt{\frac{M}{2\pi q(1-q)}}
\exp\!\left(-\frac{M(x-q)^2}{2q(1-q)}\right)\right]^{1/2}\\
&\quad\times\left[\int_{\pm A}^{\pm\infty}\!\!dq\;\sqrt{\frac{M}{2\pi q(1-q)}}
\exp\!\left(-\frac{M(x-q)^2}{2q(1-q)}\right)\right]^{1/2}.
\end{align*}
The first factor can be bounded by the square root of an expectation (under the Gaussian kernel centered at \(x\) with variance \(\sim 1/M\)) of \(|q|^{2\alpha-2}\), and is finite for our parameter range. The second factor is exponentially small in \(M\) by Mills' ratio. Hence the tail contribution is negligible for large \(M\), allowing us to write
\begin{equation}\label{eq:px_full_int}
    p(x)\propto \int_{-\infty}^{\infty} dq \; q^{\alpha-1}\,
    \sqrt{\frac{M}{2\pi q(1-q)}}\,
    \exp\!\left[-\frac{M(x-q)^2}{2q(1-q)}\right]
    +\mathcal{O}\!\left(\frac{1}{M}\right)\,.
\end{equation}
For large \(M\) the integrand in \eqref{eq:px_full_int} is sharply concentrated near \(q=x\), so we apply Laplace’s method. Define the integral (up to the overall proportionality) by
\[
I_M(x)=\int_{-\infty}^{\infty} dq\; q^{\alpha-1}\sqrt{\frac{M}{2\pi q(1-q)}}\,
\exp\!\Big[-\frac{M(x-q)^2}{2q(1-q)}\Big].
\]
Fix \(x\in(0,q_s)\) and choose \(\delta>0\) so small that \(x\in[\delta,q_s-\delta]\); then there is \(c>0\) with \(q(1-q)\ge c\) for all \(q\in[x-\delta,x+\delta]\). Introduce the slowly varying prefactor
\[
A(q):=q^{\alpha-1}\sqrt{\frac{1}{q(1-q)}},
\]
and assume \(A\) is \(C^3\) in a neighborhood of \(q=x\). Let
\[
M_j:=\sup_{q\in[x-\delta,x+\delta]}|A^{(j)}(q)|,\qquad j=0,1,2,3.
\]
Define the local width
\[
\sigma\equiv\sqrt{\frac{x(1-x)}{M}},
\]
and change variables \(q=x+\sigma u\). A direct computation yields
\[
I_M(x)=\frac{1}{\sqrt{2\pi}}\int_{-\infty}^{\infty} C(u;\sigma)
\exp\!\Big[-\tfrac{1}{2}u^2\Phi(u;\sigma)\Big]\,du,
\]
with
\[
C(u;\sigma)\equiv A(x+\sigma u)\sqrt{\frac{x(1-x)}{(x+\sigma u)(1-x-\sigma u)}},
\qquad
\Phi(u;\sigma)\equiv\frac{x(1-x)}{(x+\sigma u)(1-x-\sigma u)}.
\]
For \(|u|\le \delta/\sigma\) the functions \(C\) and \(\Phi\) are smooth in \(\sigma\) and admit Taylor expansions. In particular there exist constants \(K_1,K_2,K_3\) (depending on \(x,\delta,M_j\)) such that, for sufficiently small \(\sigma\),
\[
\big|\Phi(u;\sigma)-1\big|\le K_1\sigma|u| + K_2\sigma^2 u^2,
\]
and
\[
\big|C(u;\sigma)-C(0;0)\big|\le M_1\sigma|u| + \tfrac{1}{2}M_2\sigma^2 u^2 + \tfrac{1}{6}M_3\sigma^3|u|^3.
\]
We rewrite
\[
I_M(x)=\frac{1}{\sqrt{2\pi}}\int_{-\infty}^{\infty} C(u;\sigma)e^{-u^2/2}
\exp\!\Big[-\tfrac{1}{2}u^2(\Phi(u;\sigma)-1)\Big]\,du,
\]
and split it into \(I_M(x)=I^{(0)}+I^{(1)}\), where
\[
I^{(0)}=\frac{1}{\sqrt{2\pi}}\int C(u;\sigma)e^{-u^2/2}du,
\qquad
I^{(1)}=\frac{1}{\sqrt{2\pi}}\int C(u;\sigma)e^{-u^2/2}\big(\exp[-\tfrac{1}{2}u^2(\Phi-1)]-1\big)du.
\]
Expand \(C(u;\sigma)\) about \(u=0\) up to second order with remainder:
\[
C(u;\sigma)=C(0;0)+\sigma u\,c_1+\frac{\sigma^2 u^2}{2}c_2+R_C(u;\sigma),
\qquad |R_C(u;\sigma)|\le \frac{M_3}{6}\sigma^3|u|^3.
\]
Integrating termwise against the even Gaussian \(e^{-u^2/2}/\sqrt{2\pi}\) kills the linear term, and yields
\[
I^{(0)} = C(0;0) + \frac{\sigma^2}{2}c_2 + O(\sigma^3).
\]
For \(I^{(1)}\) use the expansion
\[
\exp\!\Big[-\tfrac{1}{2}u^2(\Phi-1)\Big]-1 = -\tfrac{1}{2}u^2(\Phi-1) + O\big(u^4(\Phi-1)^2\big),
\]
together with the bound \(|\Phi-1|\le K_1\sigma|u|+K_2\sigma^2u^2\). One obtains
\[
|I^{(1)}|\le K'\big(\sigma\!\int |u|^3 e^{-u^2/2}du + \sigma^2\!\int u^4 e^{-u^2/2}du\big),
\]
for some constant \(K'\). Since the Gaussian moments are finite,
\[
|I^{(1)}|\le K''(\sigma+\sigma^2).
\]
Odd terms coming from products of odd expansions vanish on integration, so the leading remainder behaves like \(\sigma^2\).
Combining the estimates for \(I^{(0)}\) and \(I^{(1)}\), and using \(\sigma^2=x(1-x)/M\), we deduce
\[
I_M(x)=C(0;0) + O(\sigma^2) + O(\sigma^3) = C(0;0) + O\!\left(\frac{1}{M}\right).
\]
Finally, \(C(0;0)=A(x)\sqrt{\dfrac{x(1-x)}{x(1-x)}}=A(x)=x^{\alpha-1}\), hence
\[
I_M(x)=x^{\alpha-1} + \mathcal{O}\!\left(\frac{1}{M}\right)\qquad(M\to\infty).
\]
Thus the Laplace approximation of \eqref{eq:px_full_int} yields the leading behavior \(x^{\alpha-1}\) with a controlled error $\mathcal{O}\!\left(\frac{1}{M}\right)$. Recalling that the binomial-to-Gaussian replacement earlier also introduced an error of order $\mathcal{O}\!\left(\frac{1}{M}\right)$, the total error in \(p(x)\) remains $\mathcal{O}\!\left(\frac{1}{M}\right)$. In conclusion,
\[
p(x)\propto x^{\alpha-1} + \mathcal{O}\!\left(\frac{1}{M}\right),
\qquad x\in(0,q_s),
\]
or equivalently
\[
p(x)\propto x^{\frac{\ln m}{\ln\!\left(\frac{1-q_s}{m}\right)}-1} + \mathcal{O}\!\left(\frac{1}{M}\right),
\]
as stated in the main text.

\clearpage
\section{Section D: Gaussian parent p.d.f.}
~\label{sec: Gaussian}

\begin{figure}[]
\centering
\includegraphics[width=\textwidth]{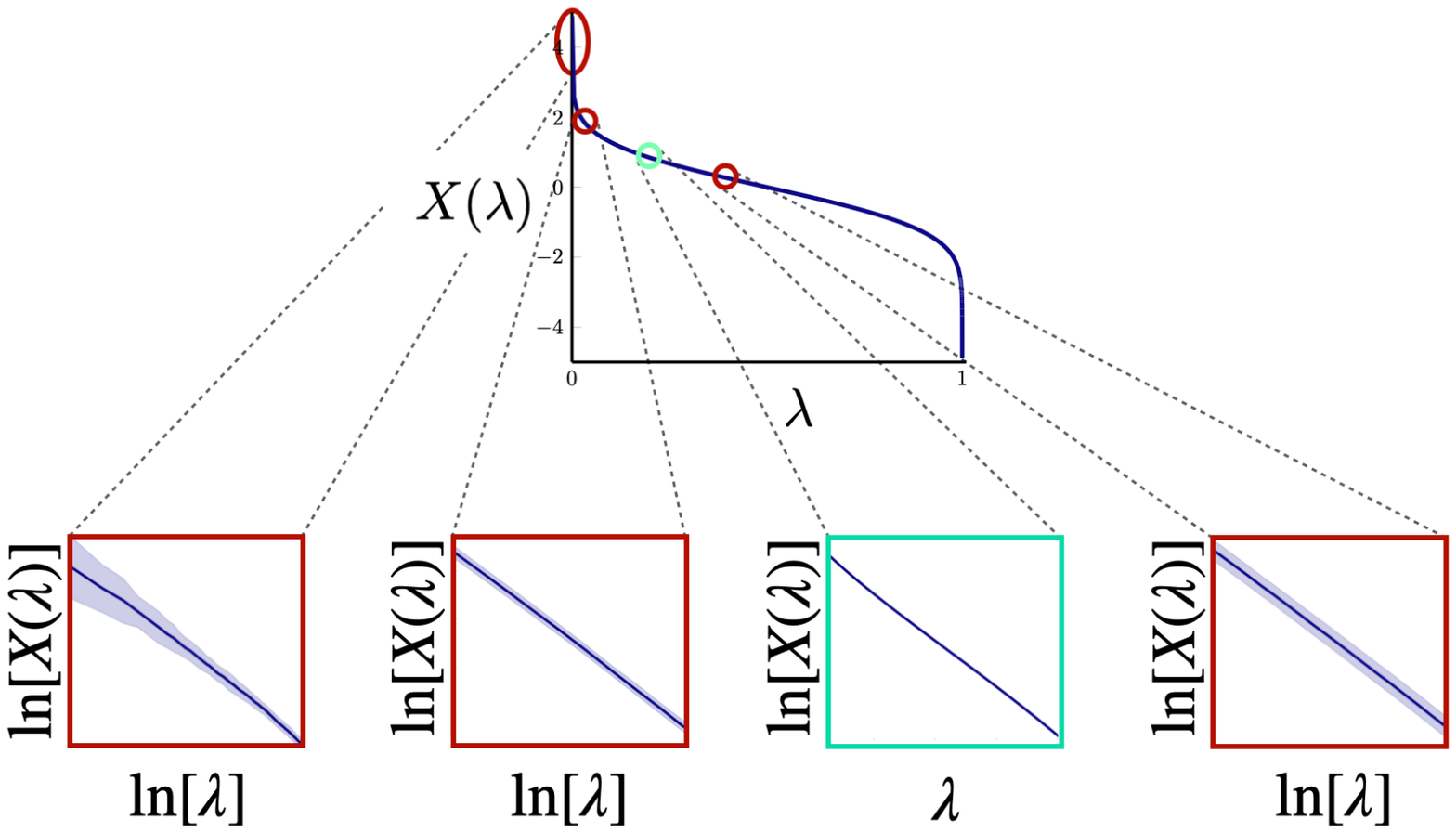}
\caption{Order statistics $X(\lambda)$ associated with a Gaussian normal parent distribution.
Numerical results are reported together with shades representing their statistical error
for a sample size of $N = 10^6$.
Zoomed regions are plotted in a log-log scale, 
where a linear behaviour clearly shows up,
highlighting the power-law nature of $X(\lambda)$. 
An only exception is made around $X(\lambda) = 1$, for which the linear behaviour is made manifest with a logy scale, confirming the expected local exponential behaviour of $X(\lambda)$.}
\end{figure}
Here we focus on the order statistics associated to a Gaussian parent p.d.f.
\begin{equation}
    p(x) = \frac{1}{\sqrt{2 \pi}\sigma}e^{-(x-\mu)^2/2\sigma^2}.
\end{equation}
In order to perform the change of variables in Eq.~\eqref{eq: change of variables}
we restrict our study to either $x \geq 0$ or $x < 0$.
In any case we define
\begin{equation}
    \begin{cases}
        \abs{x} = e^\eta\\
        p(x) = e^{z(\eta)},
    \end{cases}
\end{equation}
from where
\begin{equation}
    z(\eta) = -\frac{1}{2}\left[\frac{\left(e^{\eta}-\mu\right)^2}{\sigma^2} + \mathrm{ln}2\pi\sigma^2 \right].
\end{equation}
The local approximation of $p(x)$ with a power law 
is equivalent to consider
\begin{equation}
    z(\eta) \approx z(H_0) + z^{(1)}(H_0)(\eta - H_0).
\end{equation}
We proved that the associated order statistics will
locally follow a Zipf's law whenever $z^{(1)}(H_0) \neq -1$,
i.e. iff
\begin{equation}
    \left(\langle X(\lambda_0) \rangle -\mu \right)\langle X(\lambda_0) \rangle \neq \sigma^2.
\end{equation}
In other words, 
the only two points of the order statistics 
with an exponential local behaviour, 
instead of a Zipf's one,
are those were $\left\langle X(\lambda_0) \right\rangle = \pm 1$, i.e.
those corresponding to one standard deviation from the mean value of the Gaussian distribution.
The local behavior around any other value of $\lambda_0$ is power-like, with the exponent $\alpha(\lambda_0)$ given by Eq.~\eqref{eq: local alpha}, that in the case of a normal parent p.d.f. reduces to
\begin{equation}
    \alpha(\lambda_0) = \left[\frac{\langle X(\lambda_0)\rangle\left(\langle X(\lambda_0) \rangle -\mu \right)}{\sigma^2}-1\right]^{-1}.
\end{equation}

\end{document}